\documentclass[runningheads,a4paper,english]{llncs}[2022/01/12]
\usepackage{upquote}

\usepackage[ngerman,main=english]{babel}
\addto\extrasenglish{\languageshorthands{ngerman}\useshorthands{"}}
\usepackage{regexpatch}
\makeatletter
\edef\switcht@albion{%
  \relax\unexpanded\expandafter{\switcht@albion}%
}
\xpatchcmd*{\switcht@albion}{ \def}{\def}{}{}
\xpatchcmd{\switcht@albion}{\relax}{}{}{}
\edef\switcht@deutsch{%
  \relax\unexpanded\expandafter{\switcht@deutsch}%
}
\xpatchcmd*{\switcht@deutsch}{ \def}{\def}{}{}
\xpatchcmd{\switcht@deutsch}{\relax}{}{}{}
\edef\switcht@francais{%
  \relax\unexpanded\expandafter{\switcht@francais}%
}
\xpatchcmd*{\switcht@francais}{ \def}{\def}{}{}
\xpatchcmd{\switcht@francais}{\relax}{}{}{}
\makeatother

\usepackage[hyphens]{url}

\makeatletter
\g@addto@macro{\UrlBreaks}{\UrlOrds}
\makeatother

\usepackage[%
    rm={oldstyle=false,proportional=true},%
    sf={oldstyle=false,proportional=true},%
    tt={oldstyle=false,proportional=true,variable=false},%
    qt=false%
]{cfr-lm}

\usepackage[T1]{fontenc}

\usepackage[
  babel=true, 
  expansion=alltext,
  protrusion=alltext-nott, 
  final 
]{microtype}

\usepackage[printonlyused]{acronym}
    \newacro{AoS}{array of structures}
\newacro{AoSoA}{array of structures of arrays}
\newacro{CPI}{cycles per instruction}
\newacro{DEM}{discrete element method}
\newacro{EAM}{embedded atom method}
\newacro{FCC}{face-centered cubic}
\newacro{FN}{full neighbor-lists}
\newacro{HN}{half neighbor-lists}
\newacro{ILP}{instruction-level parallelism}
\newacro{IR}{intermediate representation}
\newacro{ISA}{instruction set architecture}
\newacro{LJ}{Lennard-Jones}
\newacro{MD}{Molecular dynamics}
\newacro{MPI}{message passing interface}
\newacro{SIMD}{single instruction, multiple data}
\newacro{SoA}{structure of arrays}
\newacro{PBC}{periodic boundary conditions}
\newacro{HPM}{hardware performance monitoring}

\DisableLigatures{encoding = T1, family = tt* }

\usepackage{graphicx}

\usepackage{diagbox}

\usepackage{xcolor}

\usepackage{listings}

\usepackage{algorithm,algpseudocode}

\definecolor{eclipseStrings}{RGB}{42,0.0,255}
\definecolor{eclipseKeywords}{RGB}{127,0,85}
\colorlet{numb}{magenta!60!black}

\lstdefinelanguage{json}{
    basicstyle=\normalfont\ttfamily,
    commentstyle=\color{eclipseStrings}, 
    stringstyle=\color{eclipseKeywords}, 
    numbers=left,
    numberstyle=\scriptsize,
    stepnumber=1,
    numbersep=8pt,
    showstringspaces=false,
    breaklines=true,
    frame=lines,
    string=[s]{"}{"},
    comment=[l]{:\ "},
    morecomment=[l]{:"},
    literate=
        *{0}{{{\color{numb}0}}}{1}
         {1}{{{\color{numb}1}}}{1}
         {2}{{{\color{numb}2}}}{1}
         {3}{{{\color{numb}3}}}{1}
         {4}{{{\color{numb}4}}}{1}
         {5}{{{\color{numb}5}}}{1}
         {6}{{{\color{numb}6}}}{1}
         {7}{{{\color{numb}7}}}{1}
         {8}{{{\color{numb}8}}}{1}
         {9}{{{\color{numb}9}}}{1}
}

\usepackage[autostyle=true]{csquotes}

\defineshorthand{"`}{\openautoquote}
\defineshorthand{"'}{\closeautoquote}

\usepackage{booktabs}

\usepackage{paralist}

\usepackage[%
  square,        
  comma,         
  numbers,       
  sort&compress 
]{natbib}

\usepackage{etoolbox}
\makeatletter
\patchcmd{\NAT@test}{\else \NAT@nm}{\else \NAT@hyper@{\NAT@nm}}{}{}
\makeatother

\SetExpansion
[ context = sloppy,
  stretch = 30,
  shrink = 60,
  step = 5 ]
{ encoding = {OT1,T1,TS1} }
{ }

\usepackage[rflt]{floatflt}

\usepackage{subfigure}
\usepackage{pdfcomment}

\usepackage{stfloats}
\fnbelowfloat

\usepackage[group-minimum-digits=4,per-mode=fraction]{siunitx}
\addto\extrasgerman{\sisetup{locale = DE}}

\usepackage{hyperref}

\hypersetup{
  hidelinks,
  colorlinks=true,
  allcolors=black,
  pdfstartview=Fit,
  breaklinks=true
}

\usepackage[all]{hypcap}

\usepackage[capitalise,nameinlink]{cleveref}

\crefname{section}{Sect.}{Sect.}
\Crefname{section}{Section}{Sections}
\crefname{listing}{List.}{List.}
\crefname{listing}{Listing}{Listings}
\Crefname{listing}{Listing}{Listings}
\crefname{lstlisting}{Listing}{Listings}
\Crefname{lstlisting}{Listing}{Listings}

\usepackage{lipsum}

\usepackage[math]{blindtext}
\usepackage{mwe}
\usepackage[realmainfile]{currfile}
\usepackage{tcolorbox}
\tcbuselibrary{listings}

\DeclareFontFamily{U}{MnSymbolC}{}
\DeclareSymbolFont{MnSyC}{U}{MnSymbolC}{m}{n}
\DeclareFontShape{U}{MnSymbolC}{m}{n}{
  <-6>    MnSymbolC5
  <6-7>   MnSymbolC6
  <7-8>   MnSymbolC7
  <8-9>   MnSymbolC8
  <9-10>  MnSymbolC9
  <10-12> MnSymbolC10
  <12->   MnSymbolC12%
}{}
\DeclareMathSymbol{\powerset}{\mathord}{MnSyC}{180}

\usepackage{xspace}

\makeatletter
\newcommand{\hydash}{\penalty\@M-\hskip\z@skip}
\makeatother

\input glyphtounicode
\graphicspath{{./figures/}}

\begin{document}

\title{MD-Bench: A generic proxy-app toolbox for state-of-the-art molecular dynamics algorithms}
\titlerunning{MD-Bench proxy-app toolbox}
\author{Rafael R. L. Machado \and Jan Eitzinger \and Harald K\"ostler \and Gerhard Wellein}
\authorrunning{Rafael Ravedutti Lucio Machado  et al.}
\institute{Friedrich-Alexander-Universität Erlangen-Nürnberg (FAU)\\Erlangen National High Performance Computing Center (NHR@FAU)}

\maketitle

\begin{abstract}
    Proxy-apps, or mini-apps, are simple self-contained benchmark codes with performance-relevant kernels extracted from real applications.
    Initially used to facilitate software-hardware co-design, they are a crucial ingredient for serious performance engineering, especially when dealing with large-scale production codes.
    MD-Bench is a new proxy-app in the area of classical short-range molecular dynamics.
    In contrast to existing proxy-apps in MD (e.g. miniMD and coMD) it does not resemble a single application code, but implements state-of-the art algorithms from multiple applications (currently LAMMPS and GROMACS).
    The MD-Bench source code is understandable, extensible and suited for teaching, benchmarking and researching MD algorithms.
    Primary design goals are transparency and simplicity, a developer is able to tinker with the source code down to the assembly level.
    This paper introduces MD-Bench, explains its design and structure, covers implemented optimization variants, and illustrates its usage on three examples.
\end{abstract}


\section{Introduction and Motivation}\label{sec:introduction}
\ac{MD} simulations are used in countless research efforts to assist the investigation and experimentation of systems at atomic level.
Both their system size and timescale are crucially limited by computing power, therefore they must be designed with performance in mind.
Several strategies to speedup such simulations exist, examples are Linked Cells, Verlet List and MxN kernels from GROMACS \cite{PALL20132641, DBLP:journals/jcc/SpoelLHGMB05}.
These improve the performance by exploiting either domain-knowledge or hardware features like SIMD capabilities and GPU accelerators.
\ac{MD} application codes can achieve a large fraction of the theoretical peak floating point performance and are therefore among the few application classes that can make use of the available compute power of modern processor architectures.
Also, cases scientists are interested in are frequently strong scaling throughput problems, a single run only exhibits limited parallelism but thousands of similar jobs need to be executed.
This fact combined with arithmetic compute power limitation makes an optimal hardware-aware implementation a critical requirement.
\ac{MD} is used in many areas of scientific computing like material science, engineering, natural science and life sciences.

Proxy-apps are stripped down versions of real applications, ideally they are self-contained, easy to build and bundled with a validated test case, which can be a single stubbed performance-critical kernel or a full self-contained small version of an application.
Historically proxy-apps assisted in porting efforts or hardware-software co-design studies.
Today they are a common ingredient of any serious performance engineering effort of large-scale application codes, especially with multiple parties involved.
Typically proxy-apps resemble a single application code, e.g., miniMD \cite{6805038} mimics the performance of LAMMPS \cite{PLIMPTON19951, BROWN2012449}.
A proxy-app can be used for teaching purposes and as a starting point for performance oriented research in a specific application domain.

To investigate the performance of MD applications, we developed MD-Bench --- a standalone proxy-app toolbox implemented in C99 that comprises the most essential \ac{MD} steps to calculate trajectories in an atomic-scale system.
MD-Bench contributes clean reference implementations of state-of-the-art MD optimization schemes.
As a result, and in contrast to existing MD proxy-apps, MD-Bench is not limited to represent one MD application but aims to cover all relevant contributions.
MD-Bench is intended to facilitate and encourage performance related research for classical MD algorithms.
Its applications are low-level code analysis and performance investigation via fine-grained profiling of hardware resources utilized by \ac{MD} runtime-intensive kernels.

This paper is structured as follows:
In \autoref{sec:relatedwork} we present related work on proxy-apps for \ac{MD} simulations, pointing out the differences compared to MD-Bench.
In \autoref{sec:md} we explain the basic theory for \ac{MD} simulations.
In \autoref{sec:features} we list and discuss current features offered in MD-Bench, besides its employment on benchmarking, performance analysis and teaching activities.
In \autoref{sec:showcases} we present cases with analysis studies and results to illustrate how MD-Bench can be used.
Finally \autoref{sec:outlook} presents the conclusion and outlook, and a discussion of future work.

\section{Related Work}
\label{sec:relatedwork}
There already exist multiple proxy-apps to investigate performance and portability for \ac{MD} applications.
One of the better known examples is Mantevo miniMD, which contains C++ code extracted from LAMMPS and provides a homogeneous copper lattice test case in which the short-range forces can be calculated with \ac{LJ} or \ac{EAM} potentials.
It was used to investigate the performance of \ac{SIMD} vectorization on the most internal loops of the neighbor-lists building and force calculation steps \cite{6569887}, as well as to evaluate the portability of \ac{MD} kernels through the Kokkos framework, thus executing most of the code on GPU instead of only the pair force and neighbor-lists.
Outcomes from miniMD in LAMMPS include better \ac{SIMD} parallelism usage on code generated by compilers and most efficient use of GPU by avoiding data transfers at all time-steps of the simulation.

ExMatEx coMD is another proxy-app from the material science domain that focuses on co-design to evaluate the performance of new architectures and programming models.
Besides allowing users to extend and/or re-implement the code as required, the co-design principle also permits to evaluate the performance when switching strategies, for example using Linked-Cells for force calculations instead of Verlet Lists.

ExaMiniMD is an improved and extended version of miniMD with enhanced modularity that also uses Kokkos for portability. Its main components such as force calculation, communication and neighbor-list construction are derived classes that access their functionality through virtual functions.

In previous work we developed tinyMD \cite{MACHADO2021101425}, a proxy-app (also based on miniMD) created to evaluate the portability of \ac{MD} applications with the AnyDSL framework.
tinyMD uses higher-order functions to abstract device iteration loops, data layouts and communication strategies, providing a domain-specific library to implement pair-wise interaction kernels that execute efficiently on multi-CPU and multi-GPU targets.
Its scope is beyond \ac{MD} applications, since it can be used to simulate any kind of particle simulation that relies on short-range force calculation such as the \ac{DEM}.

Beyond proxy-apps, performance-engineering of \ac{MD} can be achieved via auto-tuning by either running simulations as a black-box for finding optimal system and hardware specific simulation parameters \cite{doi:10.1063/5.0019045} or by providing programming interfaces that dynamically tune the application at run-time by selecting the best optimization strategies and data layouts \cite{8778280}.

MD-Bench differs from available offerings because it was primarily developed to enable an in-depth analysis of software-hardware interaction.
With that in mind, MD-Bench provides a stubbed variant to evaluate the force calculation at different levels of the memory hierarchy, as well as a gather benchmark that mimics the memory operations used in those kernels, thus allowing investigation of the memory transfers without side-effects from arithmetic operations.
In contrast to other proxy-apps MD-Bench contains optimized algorithms from multiple MD applications and allows to compare those using the same test cases.
Apart from standard Verlet Lists algorithms it contains state-of-the-art optimization strategies as, e.g., the GROMACS MxN kernels, which attain higher data level parallelism in modern architectures by using a more \ac{SIMD}-friendly data layout and up to now are not yet available as part of a simple proxy-app.
Although the significant majority of MD-Bench code is implemented in C, it also relies on SIMD intrinsics and assembly code kernels allowing low-level tweaking of the code without interference from a compiler.

\section{Background and Theory}
\label{sec:md}



Fundamentally, atom trajectories in classical \ac{MD} systems are computed by integrating Newton's second law equation (\autoref{eq:newton}) after computing the forces, which are described by the negative gradient of the potential of interest.
\autoref{eq:lennard_jones} shows how to compute each interaction for the \ac{LJ} potential, with $x_{ij}$ being the distance vector between atoms $i$ and $j$, $\epsilon$ being the width of the potential well, and $\sigma$ specifying at which distance the potential is~$0$.
\begin{equation}
    F = m \dot{v} = m a \label{eq:newton} \\
\end{equation}
\begin{equation}
    F_{2}^{LJ}(x_i, x_j) = 24\epsilon \left( \frac{\sigma}{x_{ij}} \right)^{6} \left[ 2\left(\frac{\sigma}{x_{ij}}\right)^{6} - 1\right] \frac{x_{ij}}{|x_{ij}|^{2}}
    \label{eq:lennard_jones}
\end{equation}



\begin{figure}
    \centering
    \subfigure[Verlet List]{\label{fig:verlet}\includegraphics[width=0.3\textwidth]{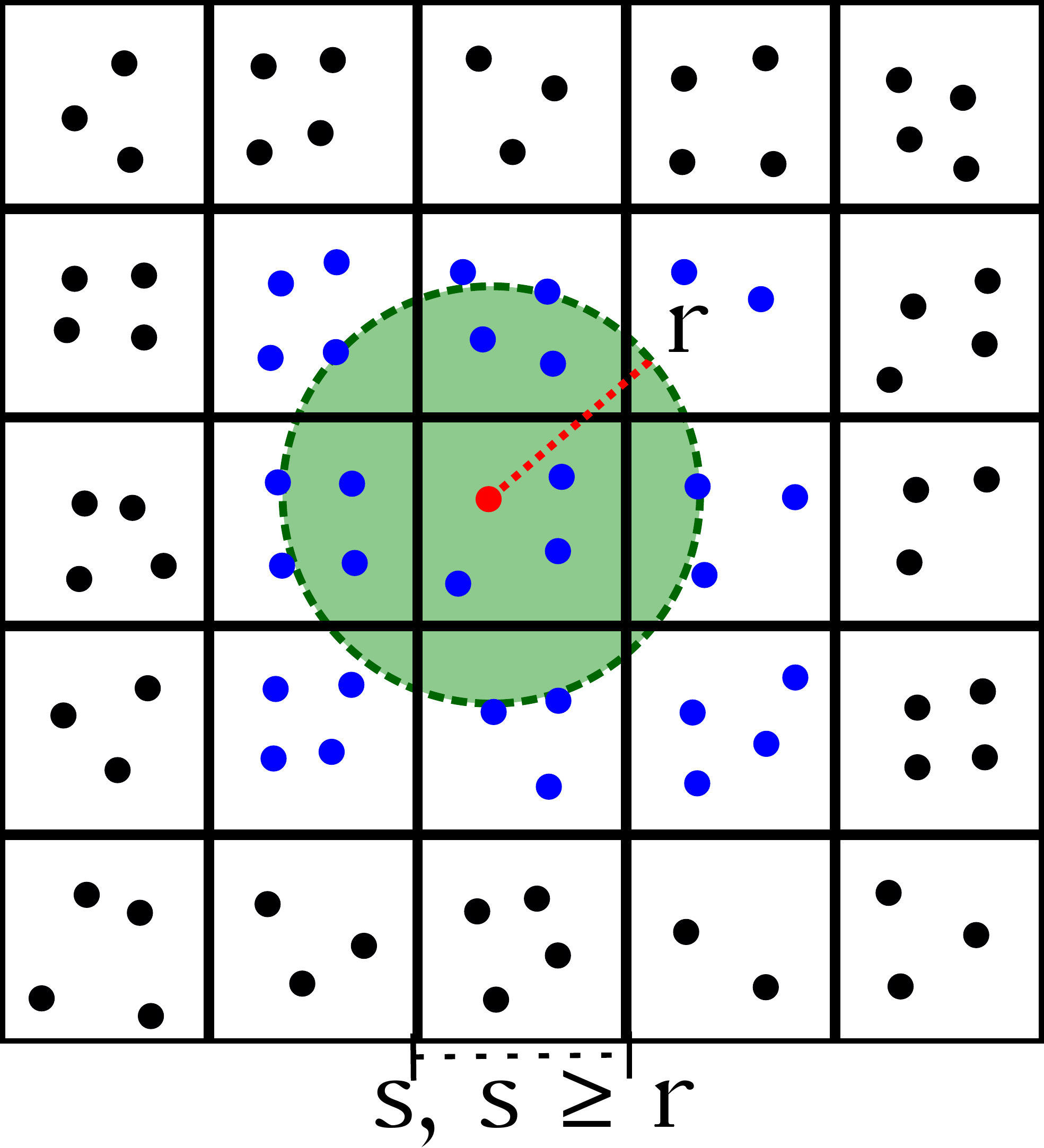}}
    \subfigure[GROMACS MxN]{\label{fig:gromacs_mxn}\includegraphics[width=0.3\textwidth]{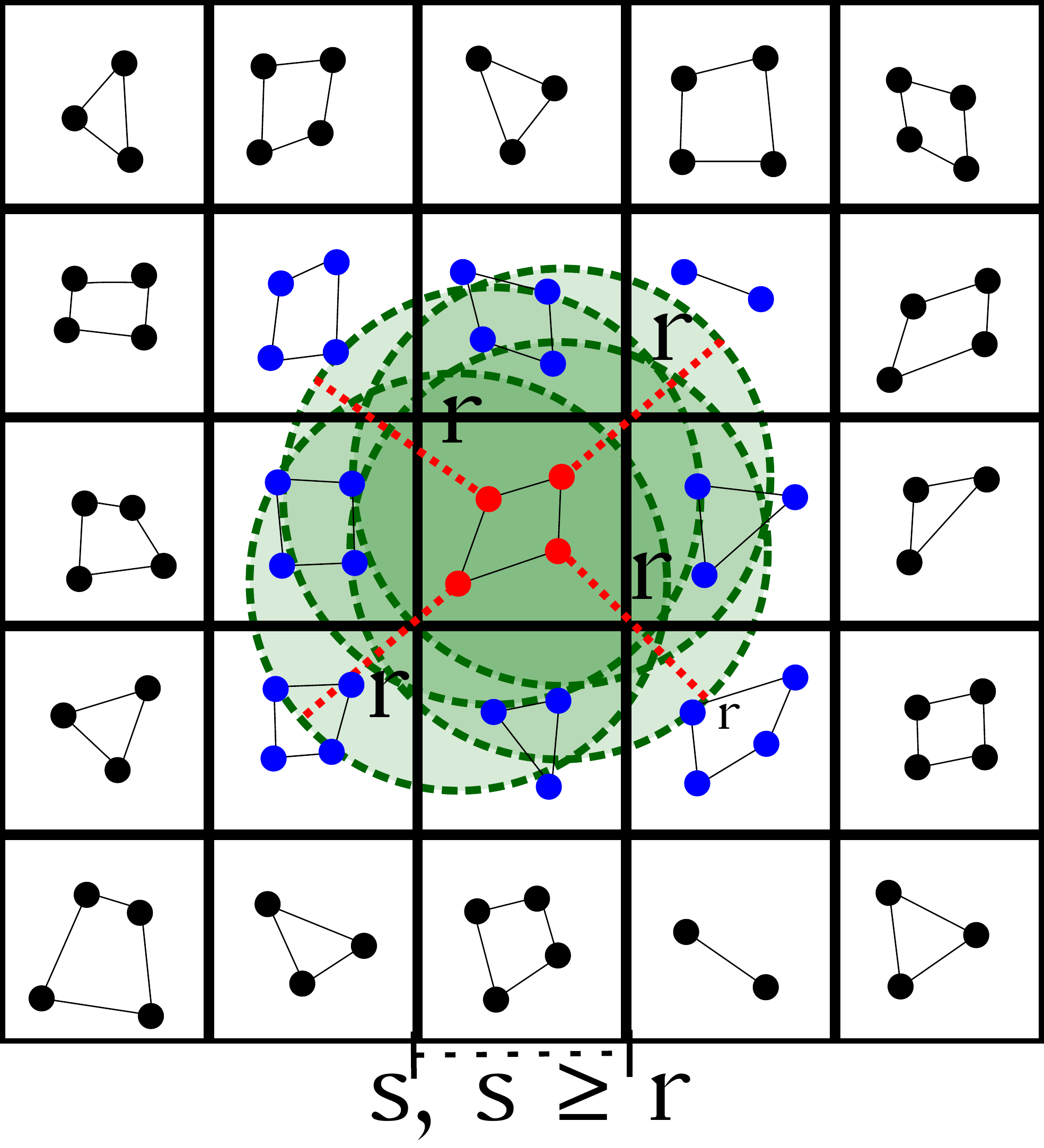}}
    \subfigure[Stubbed Patterns]{\label{fig:stub}\includegraphics[width=0.3\textwidth]{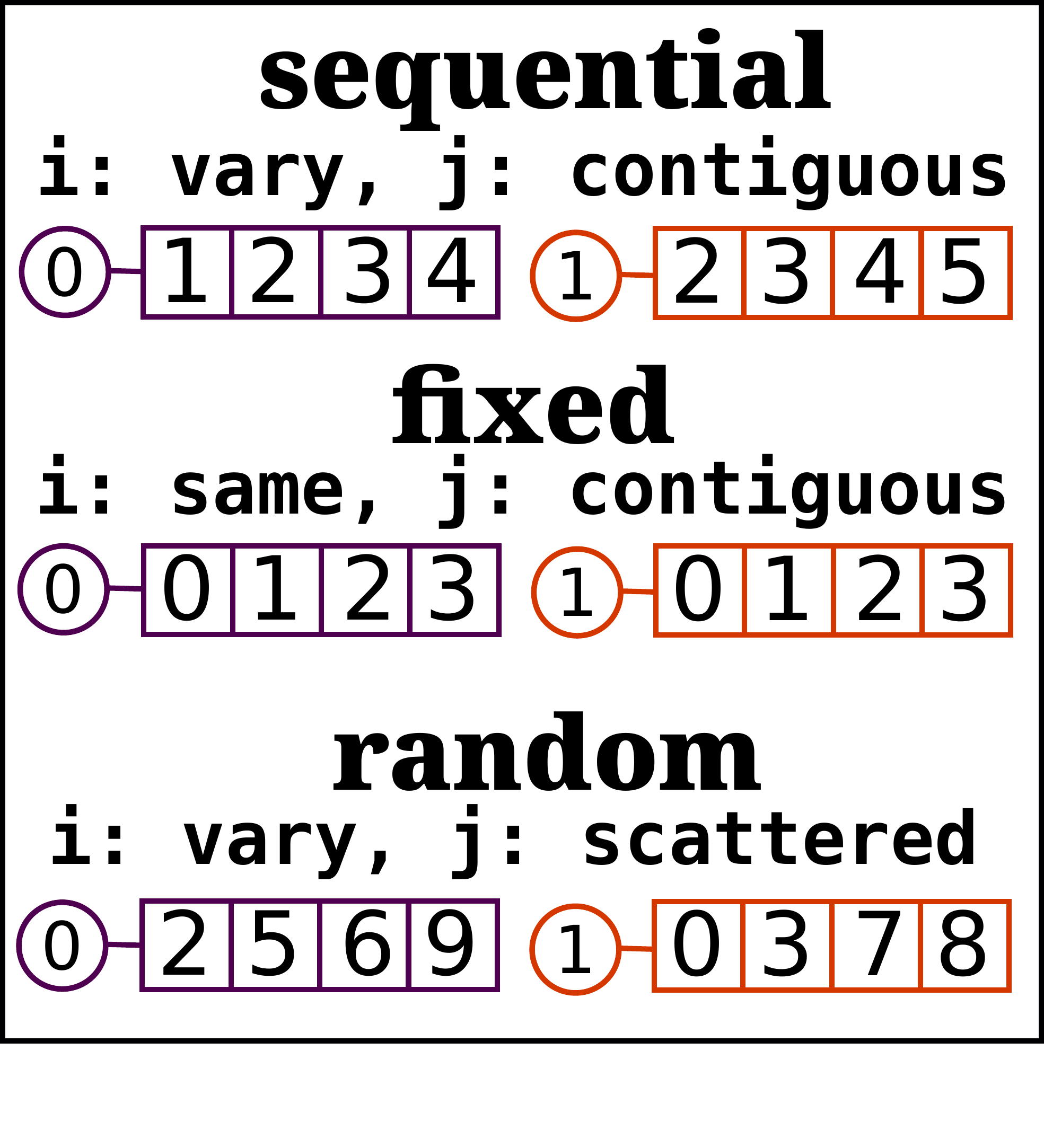}}
    \caption{(a, b) Pair list creation for the red atom (cluster) in Verlet List (GROMACS MxN), blue atoms (clusters) are evaluated and the ones within the green circle with radius $r$ are added to the pair list. Cell size $s$ must be greater or equal than $r$. (c) 4-length neighbor-lists for atoms 0 (purple) and 1 (orange) in Stubbed Case patterns.}
\end{figure}

To optimize the computation of short-range potentials, only atom pairs within a cutoff radius may be considered because contributions become negligible at long-range interactions.
Hence, a Verlet List (see Fig.~\ref{fig:verlet}) can be regularly created for each atom to track neighbor candidates within a specific radius $r$, which is the cutoff radius plus a small value (verlet buffer).

\section{MD-Bench Features}
\label{sec:features}

\begin{figure}
    \centering
    \includegraphics[width=0.9\textwidth]{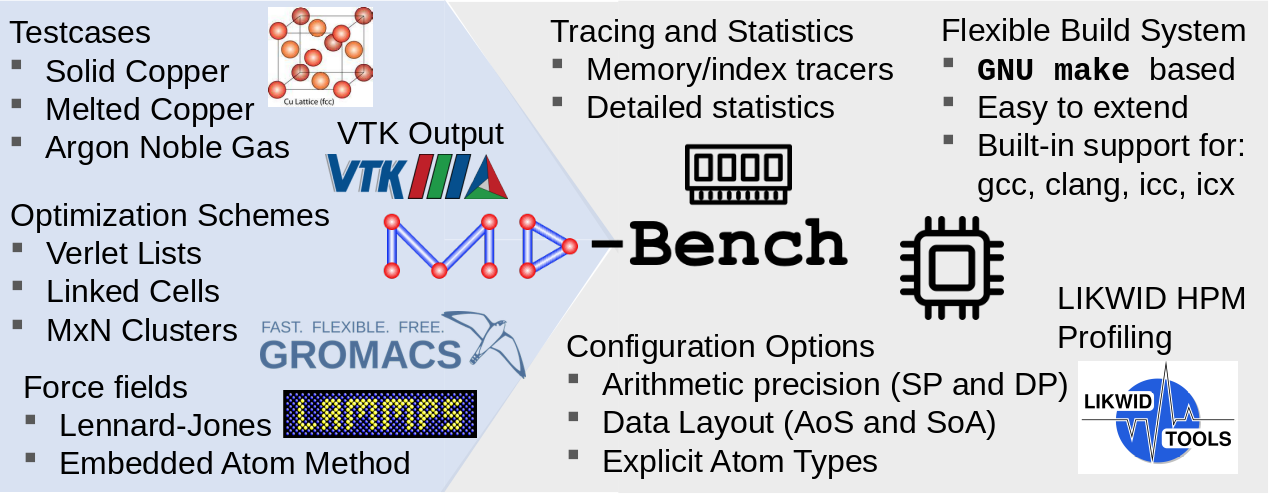}
    \caption{Overview about MD-Bench Features}
    \label{fig:features}
\end{figure}


\autoref{fig:features} depicts MD-Bench\footnote{\url{https://github.com/RRZE-HPC/MD-Bench}}$^{,}$\footnote{MD-Bench is open source and available under LGPL3 License.} features. To facilitate experimentation with a range of settings that influence performance, a robust build system with various configurations from the compiler and flags to whether atom types should be explicitly stored and loaded from memory is available.
Due to its modularity, the build system permits to replace kernels at the assembly level.
Particularly, we maintain simplified C versions of each kernel with an eye toward low-level code analysis and tweaking, which is hardly achievable on production \ac{MD} applications due to the massive code base size and extensive employment of advanced programming language techniques, which brings in substantial complexity and makes an analysis more difficult.
Apart from that, kernels are instrumented with LIKWID~\cite{psti} markers to allow fine-grained  profiling of kernels using \ac{HPM} counters.



\subsection{Optimization schemes}
\subsubsection{Verlet neighbor-lists}

The most common optimization scheme used in \ac{MD} applications to compute short-range forces is arguably the Verlet List algorithm.
It consists of building a neighbor-list for each atom in the simulation, where the elements are other atoms that lie within a cutoff radius that is higher or equal than the force cutoff radius.
Thus, forces are computed for each atom by traversing their neighbor-list and accumulating forces for neighbors which distance is smaller than the cutoff (see Algorithm \autoref{alg:force}).


\begin{algorithm}
\caption{Force calculation}
\label{alg:force}
\begin{algorithmic}
\Statex
    \scriptsize
    \For{$i \gets 1$ to $Nlocal$} \Comment{Number of local atoms}
    \State {$f$ $\gets$ {$0$}} \Comment{Partial forces (required for parallelism)}
    \For{$k \gets 1$ to $Nneighs[i]$} \Comment{Number of neighbors for atom $i$}
        \State {$j$ $\gets$ {$neighbors[i,k]$}} \Comment{$k$-th neighbor of atom $i$}
        \State {$d$ $\gets$ {$calculate\_distance(i, j)$}}
        \If {$d\leq cutoff\_radius$} \Comment{Force cutoff check}
            \State {$f$ $\gets$ {$f+calculate\_force(d)$}} \Comment{Depends on the potential used}
        \EndIf
    \EndFor
    \State {$force[i]$ $\gets$ $force[i] + f$} \Comment{Accumulate forces}
\EndFor
\end{algorithmic}
\end{algorithm}

The algorithm contains two major drawbacks with respect to SIMD parallelism: (a) irregular access pattern, since neighbor atoms are scattered across memory and (b) no reuse of neighbor data (positions) across atom iterations.
The consequence for them is the requirement of a gather operation to load the neighbor atoms data into the vector registers for each atom traversed in the outermost loop.

There are two strategies to gather data, namely \emph{hardware} or \emph{software} gathers.
Hardware gathers on x86-64 processors use a \emph{vgather} instruction
to perform the gathering at hardware level, where software gathers emulate the gather operation using separate instructions to load, shuffle and permutate elements within vector registers.
When available in the target processor, hardware gathers clearly use less instructions, but cannot take advantage of spatial locality in the \ac{AoS} layout, thus requiring at least three data transfers instead of one when keeping elements aligned to the cache line size.
Due to this trade-off between instruction execution and memory transfer, it is not straightforward to determine the best strategy, as it will depend on the target processor.

To avoid the costs of gathering data and enhance data reuse, a different data layout is necessary.
Therefore, we also introduce the GROMACS MxN optimization scheme in MD-Bench, which is currently not present in any of the existing \ac{MD} proxy-apps.
Despite its \ac{SIMD}-friendly data layout, it introduces extra computations because interactions are evaluated per clusters instead of per atom, thus the trade-off between extra computations and overhead for SIMD parallelism must also be assessed.

For all kernel variants available, MD-Bench contains their half neighbor-lists counterpart, which take advantage of Newton's Third Law to compute only half of the pair interactions, thus decreasing the amount of partial forces to calculate.
Despite the clear benefit of having less operations, this strategy harms parallelism because it introduces race conditions for the neighbor atoms because forces are stored back to memory in the innermost loop.
Also, gather and scatter operations are needed for the forces of the neighbor atoms, which not only takes effect in the instruction execution cost but also increases memory traffic.

\subsubsection{MxN cluster algorithm}
\label{sec:gromacs}

To address the lack of data reuse and necessity of costly gather operations, the MxN algorithm introduced in GROMACS clusters atoms in groups of $max(M,N)$ elements (see Figure \ref{fig:gromacs_mxn} for $M=4, N=4$ example).
Thus, tailored kernels with SIMD intrinsics are implemented to compute interactions of MxN atoms.
Positions for atoms in the same cluster are contiguously stored in an \ac{AoSoA} fashion, and can be loaded without a gather operation.
In this algorithm, $M$ parametrizes the reusability of data as atoms in the same $i$-cluster contains the same pair lists, and therefore a single load of a $j$-cluster of size $N$ is enough to compute the interaction among all pairs of atoms.
Since two kernel variants for the force calculation are present (namely \textbf{4xN} and \textbf{2xNN}), then $N$ is optimally chosen as either the SIMD width of the target processor or half of it. Since kernels must be kept simple and bidirectional mapping between $i$-clusters and $j$-clusters is needed, $M$ is either $\frac{N}{2}$, $N$ or $2N$.
Note that when a cluster size is less than $max(M,N)$, it is filled in with atoms placed in the infinity to fail the cutoff checking.
Also, different atoms in paired clusters may not be within the cutoff radius.
Hence, choosing large values for $M$ and $N$ can significantly grow the amount of pairs interactions to compute, wasting resources and injuring performance.
\subsection{Benchmark test cases}\label{sec:testcases}
Short-range force kernels for \ac{LJ} and \ac{EAM} potentials are available, with Copper \ac{FCC} lattice and Noble gases (pure argon) setups to embrace both material modeling and bio-sciences simulation fields.
Setups are provided by providing atoms positions (and velocities in some cases) in a Protein Data Bank (PDB), Gromos87 or LAMMPS dump file.

\subsection{Tools}\label{sec:tools}
\textbf{Detailed statistics mode} To collect extended runtime information for the simulation, a detailed statistics mode can be enabled.
It introduces various statistics counters in the force kernels in order to display relevant performance metrics such as cycles per SIMD iterations (processor frequency must be fixed), average cutoff conditions that fail/succeed and useful read data volumes.

\textbf{Memory traces and gather-bench} The gather benchmark is a standalone benchmark code for x86-64 CPUs (currently) that mimics the data movement (both operations and transfers) from \ac{MD} kernels in order to evaluate the ``cost of gather'' from distinct architectures.
It currently gathers data in the following patterns: (a) simple 1D arrays with fixed stride, to evaluate single gather instruction in the target CPU, (b) array of 3D vectors with fixed stride, to evaluate \ac{MD} gathers with regular data accesses and (c) array of 3D vectors using trace files from MD-Bench \emph{INDEX\_TRACER} option, to evaluate \ac{MD} gathers with irregular data accesses.
The benchmark exploits the memory hierarchy by adjusting the data volume to fit into a different cache level in successive executions, and yields a performance metric based on the number of cache lines touched.
There are options to determine the floating point precision and include padding to assure alignment to the cache line size.

\textbf{Stubbed force calculation}\label{sec:stubbed} To execute \ac{MD} kernels in a steady-state and understand their performance characteristics, we established a synthetical stubbed case where the number of neighbors per atom remain fixed and the data access pattern can be predicted.
Fig.~\ref{fig:stub} depicts examples for data access patterns available (namely sequential, fixed and random), indicating whether neighbor-lists vary or are the same across different atoms in the outermost loops (i) and if atoms in the lists (j) are contiguous or scattered over memory.
It addresses both irregular data accesses and variations in the inner-most loop size, thus providing a stable benchmark that can isolate effects caused by distinct reasons like memory latency and overhead for filling in the CPU instruction pipeline.


\section{Examples}\label{sec:showcases}
For this work experiments, we made use of the following processor architectures:
\begin{description}
    \item[Intel Cascade Lake:] Intel(R) Xeon(R) Gold 6248 CPU at 2.50GHz, with two 20-cores sockets, two threads per core (hyper-threading enabled). Individual 32KB L1 and 1024KB L2 caches for each core, and 28MB of shared L3 cache for each socket, with one memory domain per socket.
    \item[Intel Ice Lake:] Intel(R) Xeon(R) Platinum 8360Y CPU at 2.40GHz, with 36-cores per chip. Individual 80KB L1 per core (32 instructions + 48 data), 512KB L2 caches for each core, and 54MB of shared L3 cache per chip.
\end{description}
We used LIKWID (V5.2) to fix the CPU frequency, pin tasks to specific cores, enable/disable prefetchers and make use of \ac{HPM} counters.

\subsection{Assembly analysis}\label{sec:asm}
This example illustrates how MD-Bench allows to evaluate the optimizations performed by compilers at the instruction code level to pinpoint possible improvements in the compiler generated code.
In the AVX512 generated code from the Intel compiler, we notice \emph{lea} and \emph{mov} instructions before gathering the data (see \autoref{lst:leaandmovs}).
Removing such instructions does not change the semantics of the code because the destination registers of the \emph{mov} operations are not read afterwards, so we can exclude them to improve performance.
For the Copper \ac{FCC} lattice case with 200 time-steps, the runtime on Cascade Lake for the force calculation is about 4.43 seconds without these instructions and about 4.81 with them, leading to a 8\% speedup.
\definecolor{cssgreen}{rgb}{0.0, 0.5, 0.0}
\lstset{
   resetmargins=true,
   xleftmargin=0cm,
   xrightmargin=0cm,
   escapechar=!,
   basicstyle=\tiny,
   commentstyle=\ttfamily\tiny\color{cssgreen}
}

\begin{minipage}{.48\textwidth}
    \setlength\fboxsep{0pt}
    \begin{lstlisting}[language={[x86masm]Assembler},
                       label={lst:leaandmovs},
                       caption={LEA and MOV instructions.}]
    ; ymm3 <- neighs[k] * 3
    vmovdqu ymm3, [r13+rbx*4]
    vpaddd ymm4, ymm3, ymm3
    vpaddd ymm3, ymm3, ymm4
    !\colorbox{red!50}{mov r10d, [r13+rbx*4]}! ; neighs[k]
    !\colorbox{red!50}{mov r9d, [4+r13+rbx*4]}! ; neighs[k+1]
    !\colorbox{red!50}{lea r10d, [r10+r10*2]}! ; neighs[k] * 3
    !\colorbox{red!50}{lea r9d, [r9+r9*2]}! ; neighs[k+1] * 3
    !\dots! ; Same for k+2, k+3,!\dots!k+7
    vgatherdpd zmm4{k1}, [16+rdi+ymm3*8]
    vgatherdpd zmm17{k2}, [8+rdi+ymm3*8]
    vgatherdpd zmm18{k3}, [rdi+ymm3*8]
    \end{lstlisting}
\end{minipage}
\hfill
\begin{minipage}{.48\textwidth}
    \setlength\fboxsep{0pt}
    \begin{lstlisting}[language={[x86masm]Assembler},
                       label={lst:corrections},
                       caption={Correction instructions.}]
    vrcp14pd  zmm24, zmm25
    vcmppd    k2, zmm25, zmm14, 1
    !\colorbox{red!50}{vfpclasspd k0, zmm24, 30}!
    kmovw     edi, k2
    !\colorbox{red!50}{knotw     k1, k0}!
    !\colorbox{red!50}{vmovaps   zmm17, zmm25}!
    and       r10d, edi
    !\colorbox{red!50}{vfnmadd213pd zmm17, zmm24, \dots}!
    kmovw     k3, r10d
    !\colorbox{red!50}{vmulpd zmm18, zmm17, zmm17}!
    !\colorbox{red!50}{vfmadd213pd zmm24\{k1\}, zmm17, zmm24}!
    !\colorbox{red!50}{vfmadd213pd zmm24\{k1\}, zmm18, zmm24}!
    \end{lstlisting}
\end{minipage}

Further instructions are also included to perform corrections after computing the reciprocal (see \autoref{lst:corrections}), which are not present in kernels with explicit SIMD intrinsics (like MxN kernels).
Their usefulness is arguable because other factors can change the results at this precision, like the order for partial forces calculation that varies significantly across different optimization and parallelization strategies.
\autoref{tab:ccmp} shows performance, temperature and pressure for the Verlet Lists algorithm with and without such instructions, as well as for the MxN algorithm.
Without them, we can perceive a performance improvement of about 11\% in the force calculation runtime.
In terms of accuracy, not only the differences for temperature and pressure are small, but even smaller than when comparing to the MxN algorithm.
\begin{table}
\centering
\label{tab:ccmp}
\scalebox{0.9}{
\begin{tabular}{|c|c|c|c|c|}
    \hline
    Algorithm & Temperature & Pressure & Time(s) \\
    \hline
    Verlet+C & $7.961495\times10^{-1}$ & $6.721043\times10^{-1}$ & 4.78 \\
    \hline
    Verlet-C & $7.961635\times10^{-1}$ & $6.721161\times10^{-1}$ & 4.27 \\
    \hline
    MxN & $7.961966\times10^{-1}$ & $6.721441\times10^{-1}$ & 3.13 \\
    \hline
\end{tabular}}
    \caption{Temperatures, pressures and runtimes for Verlet with/without corrections (+C/-C) and MxN. Quantities are unitless and reflect \emph{lj} style from LAMMPS.}
\end{table}

\subsection{Investigate memory latency contributions}
\label{sec:latency}
An important assumption for performance engineering of streaming kernels is their non-significant latency contribution due to regular data access pattern, which is trivially foreseeable by cache prefetchers.
On a first thought, \ac{MD} simulations are expected to have significant latency impact due to their memory access characteristics, and MD-Bench can assist on measuring such impact via its stubbed sequential case.
Nonetheless, when executing them against standard copper lattice and melting cases, we observed that such impact is minor.
Furthermore, we also compare measurements in Cascade Lake from our stubbed version with kernel throughput predictions under ideal conditions from IACA \cite{iaca} (for Skylake-X micro-architecture), OSACA \cite{8641578} and uiCA \cite{Abel22} static code analyzers.

\begin{figure}
    \centering
    \includegraphics[width=8cm]{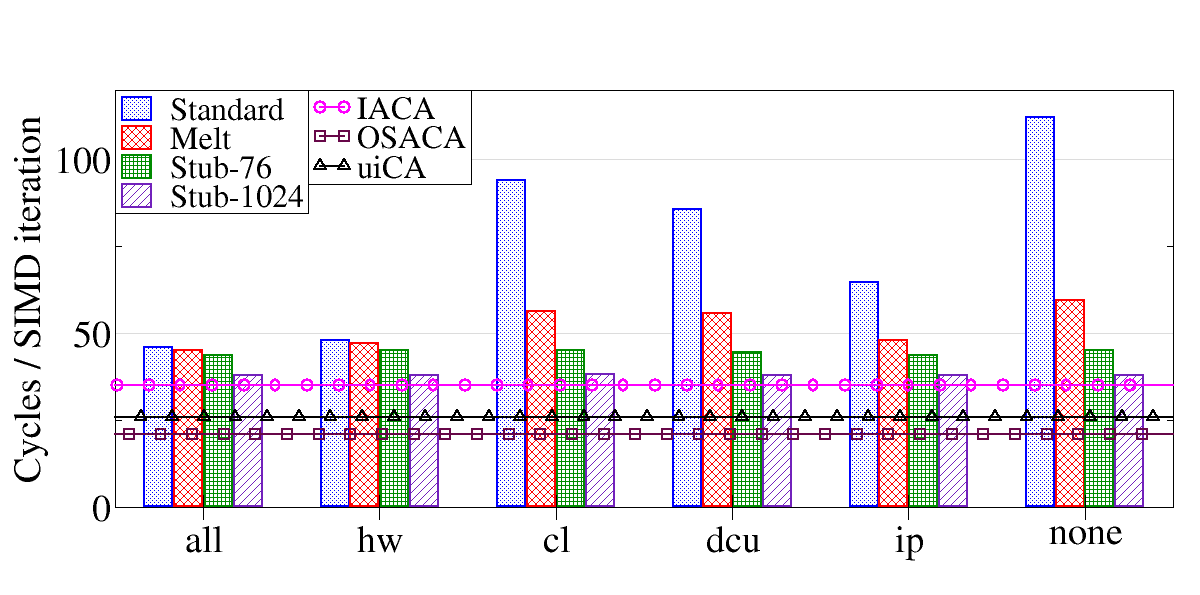}
    \caption{Cycles per SIMD iterations on Standard, Melting and Stubbed setups with different prefetchers enabled, together with predictions from static analyzers.}
    \label{fig:latency}
\end{figure}

Fig.~\ref{fig:latency} depicts cycles per SIMD iteration for mentioned cases with distinct prefetcher settings, together with IACA, OSACA and uiCA predictions.
For the stubbed case, the impact for disabling prefetchers is negligible as expected, and two versions with different number of neighbors per atom (76 and 1024) are shown to evaluate the overhead contribution from control flow divergence.
With unsteadier memory accesses, the average cycles grows by 2.1 (4\%) in standard case and 1.3 (3\%) in melting case with all prefetchers, a similar behavior with only the hardware prefetcher enabled, which makes it the most effective one.
From 76 to 1024 neighbors per atom, the number of cycles decreases by about 5.8 (15\%), hence control flow divergence contribution is higher than latency contribution.
Predictions from OSACA and uiCA are too optimistic with only 55\% and 68\% of best execution, respectively, where IACA prediction matches 92\%.
IACA reports stalled backend allocation in the CPU due to frontend bubbles, but a frontend model is not present in OSACA and uiCA.
Frontend stalls therefore contribute significantly to kernel throughput, and based on IACA results our stubbed case is close to optimal execution.

\subsection{Compiler code quality study} \label{sec:codequality}
\begin{figure}
    \centering
    \subfigure[Runtime]{\label{fig:study_runtime}\includegraphics[width=0.32\textwidth]{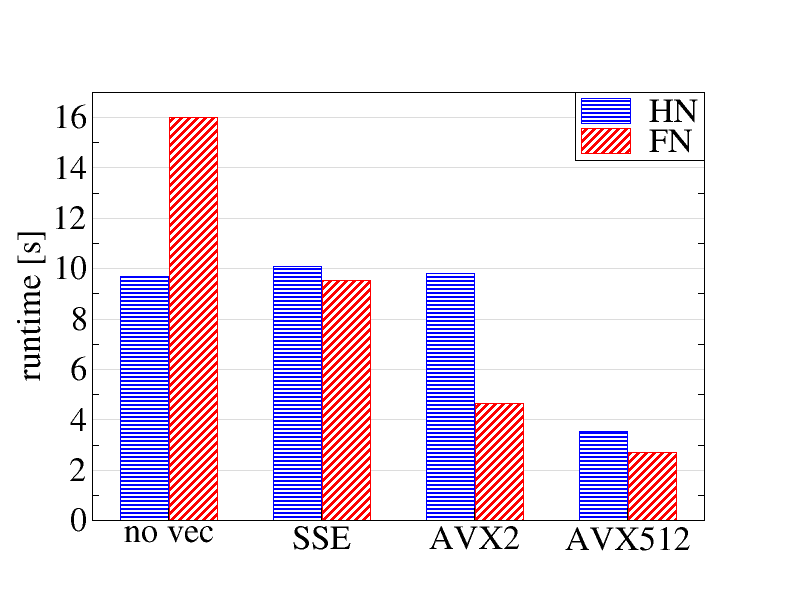}}
    \subfigure[HN Performance Profile]{\label{fig:study_hn_profile}\includegraphics[width=0.32\textwidth]{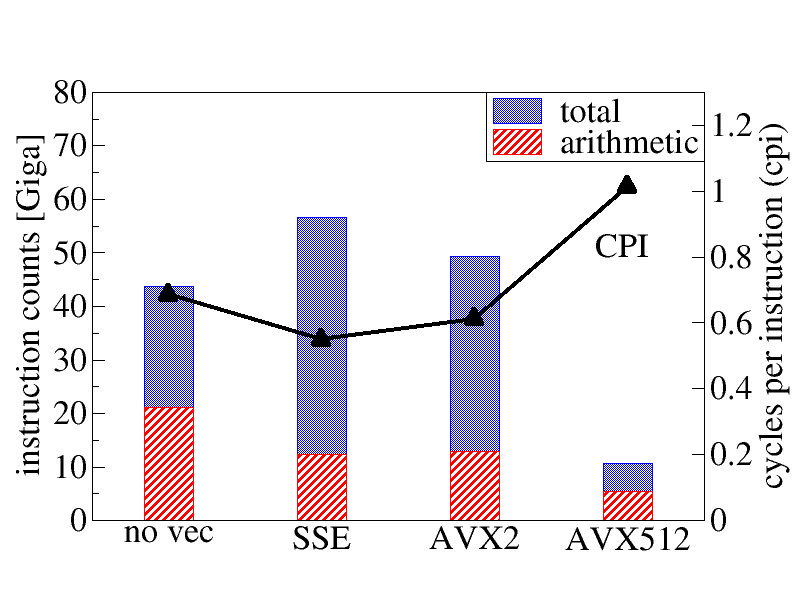}}
    \subfigure[FN Performance Profile]{\label{fig:study_fn_profile}\includegraphics[width=0.32\textwidth]{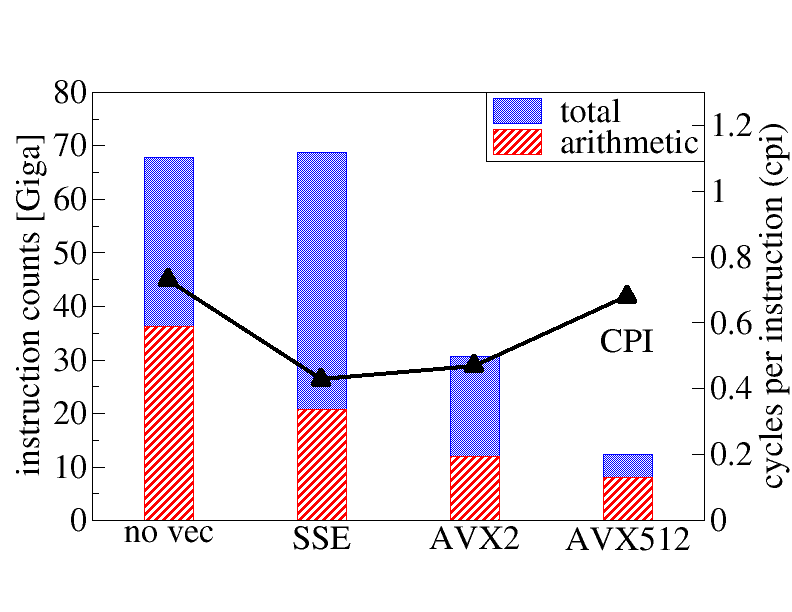}}
    \caption{Runtime (subfigure (a)), and HPM counter profiling results (Half neighbor-list HN subfigure (b) and full neighbor-list FN subfigure (c)) for the Lennard-Jones copper lattice testcase. Results are shown using compiler flags to enforce no SIMD vectorization, SSE (16b), AVX2 (32b), and AVX512 (64b) SIMD vectorization. In subfigures (b) and (c) the stacked bars show total and arithmetic instruction counts on the left y-axis, the black lines cycles per instruction (CPI) on the right y-axis.}
\end{figure}
In this example the Verlet List algorithm with \ac{HN} and \ac{FN} is benchmarked and profiled to analyze how well these are suited for vectorization and to explain the observed runtimes in more detail.
\ac{AoS} data layout was used with double precision floating point arithmetic.
Please note that \ac{HN} in MD-Bench currently does not implement the Ghost Newton optimization, hence its benefits are not accounted for.
This study was performed on the Intel Ice Lake node using the Intel compiler (ICC) 2021.4.0.
The force field kernel was compiled for several target \ac{SIMD} instruction sets and without vectorization using the \verb+-no-vec+ option.
The compiler requires a \verb+#pragma omp simd+ to vectorize the \ac{HN} variant.
The binaries were benchmarked with Turbo mode enabled, all executions were performed in the same cluster node with the same frequency, which was endorsed via HPM measurements.

Fig.~\autoref{fig:study_runtime} shows the runtime for the standard Copper lattice test case.
Without vectorization, the \ac{HN} variant is as expected faster by almost a factor of two.
When using wider \ac{SIMD} units, \ac{FN} shows almost linear speed-up and is faster than \ac{HN} for all \ac{SIMD} widths.
\ac{HN} gets slower for SSE, stagnates for AVX2, and then improves by a large step with AVX512 but still being  23\% slower than \ac{FN}.
For instruction throughput bound codes, the best case uses least instructions combined with optimal pipelined and superscalar execution, improving \ac{ILP} in the processor hardware used.
Both aspects can be directly measured using \ac{HPM} counters.
Fig.~\autoref{fig:study_hn_profile} and Fig.~\autoref{fig:study_fn_profile} show instruction counts and \ac{CPI} measurements for all \ac{HN} and \ac{FN} variants.
For \ac{HN} with SSE it can be seen that the arithmetic instruction count is almost half due to using the SSE 16b registers.
Still, the compiler does not manage to reduce the overall instruction count.
29.6\% more instructions are required to get the operands into the \ac{SIMD} registers.
The additional instruction work is partially compensated by an improved \ac{CPI} resulting in only 3.7\% worse runtime.
An explanation for this improved \ac{CPI} is that the register/register SIMD instructions on Intel processors are executed on different scheduler ports than the arithmetic instructions and therefore can be executed out-of-order.
The compiler refused to employ 32b arithmetic \ac{SIMD} instructions in the AVX2 variant, the instruction count was still decreased and the runtime slightly improved.
The enhanced capabilities of the AVX512 instruction set extension enables the compiler to generate a version with just 25\% of the instruction count of the no-vec variant.
The runtime advantage is smaller because this instruction mix is executed with a significantly worse \ac{CPI} of $1.01$.
It is still impressive that a code that was impossible to vectorize efficiently with the previous \ac{SIMD} instruction set extensions now shows an instruction count reduction of a factor of four (out of the optimal eight).

For FN the compiler manages to reduce the arithmetic instruction count with every wider \ac{SIMD} unit.
The overall instruction count increased slighly for SSE, but then is just 45\% for AVX2 and 18\% for AVX512 compared to the no-vec variant.
This underscores that the FN version is very well suited for SIMD vectorization.
This kind of study gives interesting insights and is easy to perform.
Apart from comparing algorithmic variants it can be applied to different processor architectures focusing on the \ac{CPI} metric or different compilers focusing on instruction counts.
While this type of study can also be done on other proxy-apps or application it is especially easy in MD-Bench, because all kernels are already instrumented for LIKWID.

\section{Conclusion and Outlook}\label{sec:outlook}
This paper introduced MD-Bench, a proxy-app toolbox for performance research of \ac{MD} algorithms.
It facilitates and encourages performance related research and provides clean implementations of state-of-the-art \ac{MD} optimization schemes such as Verlet List and GROMACS MxN.
We list and describe the most important MD-Bench features and its differences to other offerings, highlighting its usage on low-level code analysis and investigation of performance implications through profiling with \ac{HPM}.
Further, we support our statements on the applicability of MD-Bench by providing three use case examples that expose interesting insights concerning the performance of short-range classical \ac{MD} kernels.

MD-Bench is mature and usable, but there are still multiple open points.
We want to consider more \ac{MD} applications as, e.g., NAMD.
MD-Bench is currently a CPU-only application with a few OpenMP parallel loops.
A competitive distributed memory parallelization based on MPI is one of the next work packages.
Another sorely missing part are implementations and specific optimization schemes for GPU accelerators.
Work on supporting GPUs has already started but is in an early stage.
We hope to encourage others with this paper to participate and contribute to the development of MD-Bench.
A project like MD-Bench is an ongoing effort keeping track with recent developments and supporting novel hardware architectures.

\subsubsection*{Acknowledgments}

The authors gratefully acknowledge the scientific support and HPC resources provided by the Erlangen National High Performance Computing Center (NHR@FAU) of the Friedrich-Alexander-Universität Erlangen-Nürnberg (FAU).
NHR funding is provided by federal and Bavarian state authorities.
NHR@FAU hardware is partially funded by the German Research Foundation (DFG) – 440719683.



\renewcommand{\bibsection}{\section*{References}} 
\bibliographystyle{splncsnat}
\begingroup
  \microtypecontext{expansion=sloppy}
  \small 
  \bibliography{paper}
\endgroup

\ \\
All links were last followed on October 5, 2020.



\end{document}